\documentclass[12pt]{iopart}

\usepackage{graphicx}
\usepackage[width=0.95\linewidth,font=small]{caption}
\usepackage{dcolumn}
\usepackage{bm}
\usepackage{xcolor} 
\usepackage{subfigure} 
\usepackage[draft]{todonotes}   

\usepackage{cite}
\graphicspath{}

\begin{document}
\title[]{Towards phonon routing:\\
Controlling propagating acoustic waves in the quantum regime}

\author{M K Ekstr\"om$^1$, T Aref$^1$, A Ask$^1$, G Andersson$^1$, B Suri$^{1,2}$, H Sanada$^{1,3}$, G Johansson$^1$ and P Delsing$^1$}

\address{$^1$Chalmers University of Technology, Microtechnology and Nanoscience, 412 96 G\"oteborg, Sweden}
\eads{\mailto{per.delsing@chalmers.se}, \mailto{maria.ekstrom@chalmers.se}}
\address{$^2$Indian Institute of Science, Department of Instrumentation and Applied Physics, Bangalore 560012, India}
\address{$^3$NTT Basic Research Laboratories, 3-1, Morinosato-Wakamiya, Atsugi, Kanagawa, 243-0198 Japan}

\begin{abstract}
We explore routing of propagating phonons in analogy with previous experiments on photons.  Surface acoustic waves (SAWs) in the microwave regime are scattered by a superconducting transmon qubit. The transmon can be tuned on or off resonance with the incident SAW field using an external magnetic field or the Autler-Townes effect, and thus the reflection and transmission of the SAW field can be controlled in time. We observe 80\,\% extinction in the transmission of the low power continuous signal and a 40\,ns rise time of the router. The slow propagation speed of SAWs on solid surfaces allows for in-flight manipulations of the propagating phonons. The ability to route short, 100\,ns, pulses enables new functionality, for instance to catch an acoustic phonon between two qubits and then release it in a controlled direction. 
\end{abstract}



\section{Introduction}
Circuit Quantum ElectroDynamics (QED) based on superconducting circuits is a common platform to investigate strong coupling between light and matter, where both cavity and waveguide based systems have shown functionality in terms of quantum information processing \cite{Schoelkopf2008,Roy2017,Gu2017}. In waveguide QED, a single atom in an open system interacts with a weak resonant propagating field such that the scattered and the incident fields interfere destructively, giving rise to extinction of the forward propagating field. In early experiments on atoms and molecules demonstrating this effect, the extinction did not exceed 12\,\% \cite{Tey2008,Wrigge2008} due to the spatial mode mismatch between the incident and scattered field of the single atom in three-dimensional space. By confining the propagating fields in a one-dimensional open transmission line and using artificial atoms based on superconducting circuits, extinctions well above 90\,\% \cite{Astafiev2010,Abdumalikov2010} were demonstrated using flux qubits, and above 99\,\% extinction was achieved \cite{Hoi2011} with transmon qubits\cite{Koch2007}. 
It was also shown that the scattered and forward propagating fields could be controlled in time using fast microwave pulses to Autler-Townes split the energy levels of the artificial atom and in that way tune the artificial atom out of resonance with the propagating photons. This was used to create a single-photon router with a maximum on-off ratio of 99\,\% and 10\,ns operation time \cite{Hoi2011}. 

The fast speed of propagating photons in quantum optics experiments can be a limiting factor for some application such as in-flight manipulation. Surface acoustic waves (SAWs), on the other hand, offer clear advantages in this respect due to their much slower propagation speed. In this paper, we propose and demonstrate in-flight manipulation of SAWs at the  single-phonon level, where the manipulation is done using a superconducting transmon qubit coupled to SAWs.  

A number of experiments with SAWs in the quantum regime have been performed. For instance, phonon assisted tunnelling of SAW-irradiated quantum dots has been demonstrated \cite{Naber2006SAWDQD,Santos2010}, and on-chip transport of electrons propelled by SAWs has been shown \cite{Barnes2000,Hermelin2011,McNeil2011}. The interaction between SAWs and a superconducting qubit at the quantum level was suggested \cite{MVG2012} and shown \cite{MVG} by Gustafsson \textit{et al}. This was achieved by placing a transmon qubit on a piezoelectric substrate and interfacing it with a SAW transducer. In later experiments, SAW resonators coupled to superconducting qubits have been demonstrated and strong coupling has been shown \cite{Manenti2017,Noguchi2017}.  Furthermore, bulk acoustic waves have been used to study atom-phonon interaction \cite{Yiwen2017,Moores2018,Satzinger2018}. SAWs might also provide a solution to connect distant qubits, acting as a quantum bus in quantum computers \cite{Schuetz2015}.
These acoustic systems open up for new possibilities to reach interesting regimes that are not feasible with other systems, due to the slow speed of the SAWs and the possibility for strong coupling to artificial atoms. For instance, the concept of giant artificial atoms leads to new physics and non-Markovian effects \cite{Kockum2014,Kockum2017,Guo2017,Andersson2018,Ask2019}.

\begin{figure*}
	\begin{minipage}[t]{0.49\textwidth}
		\vspace{0pt}
		\centering
		\includegraphics[width=1\textwidth]{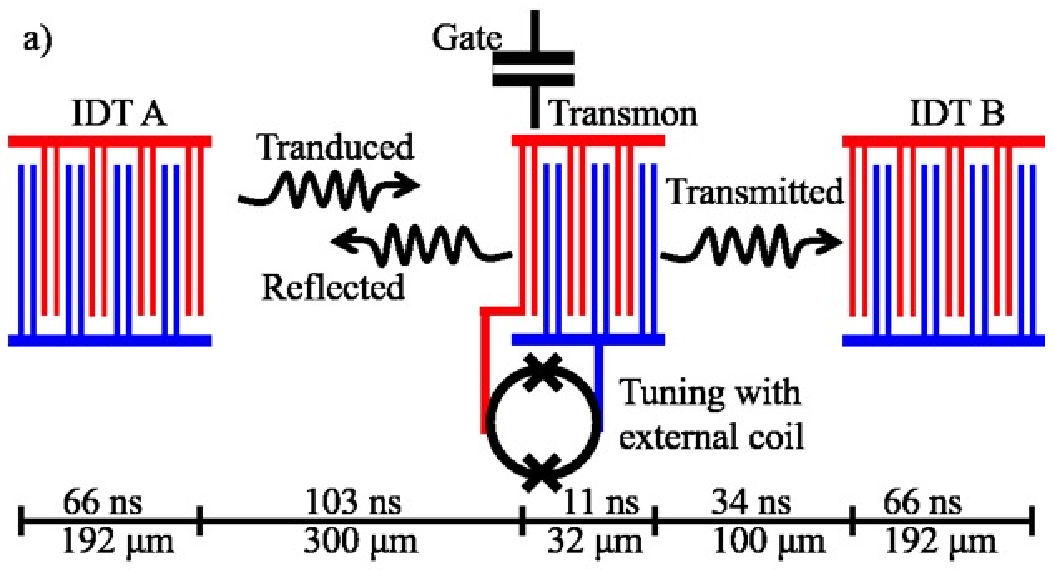}
		\label{fig:fig1a}
		\begin{minipage}[t]{1\textwidth}
			\vspace{-1cm}
			\includegraphics[width=0.63\textwidth]{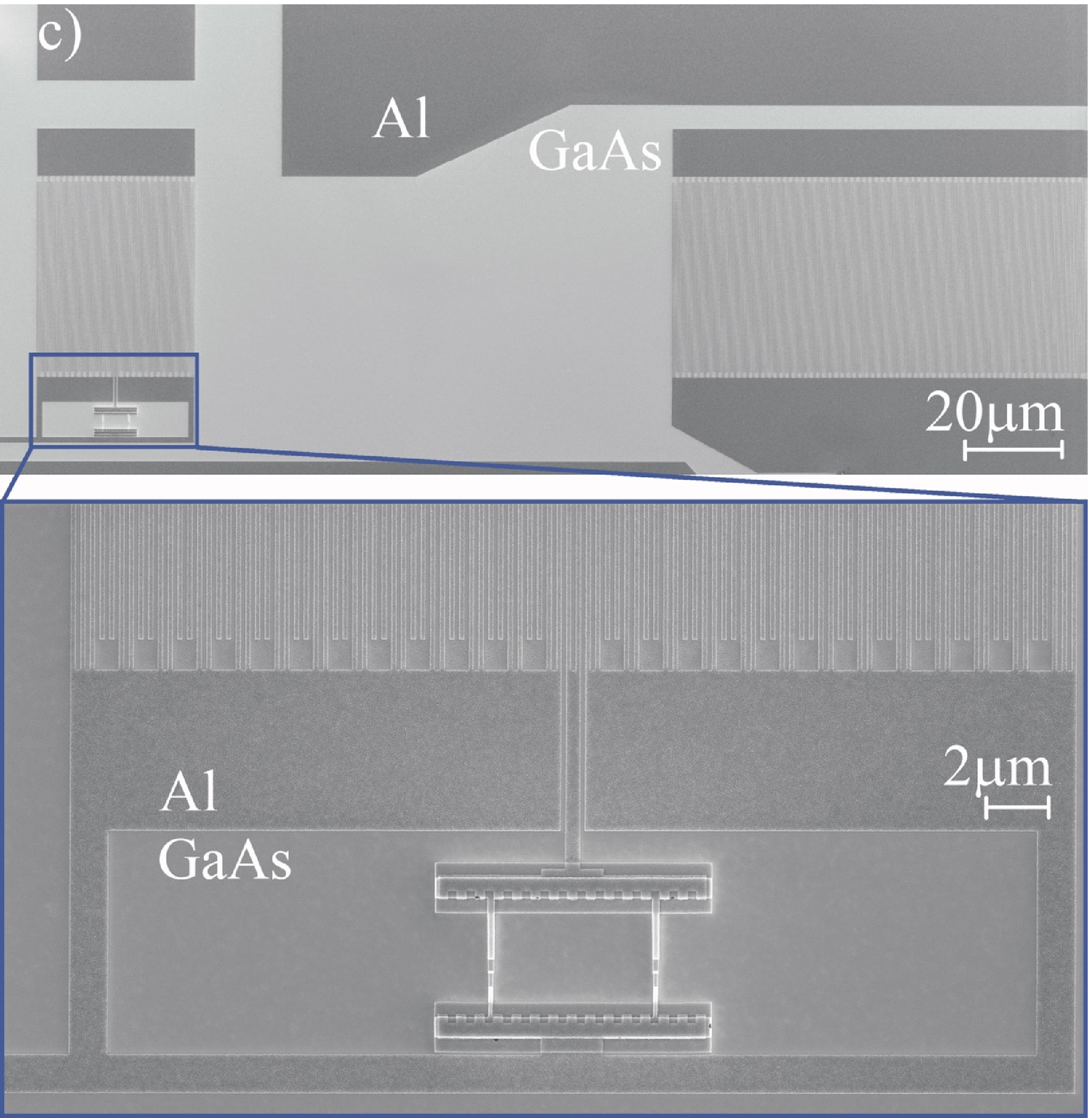} \label{fig:fig1c}
  			\includegraphics[width=0.35\textwidth]{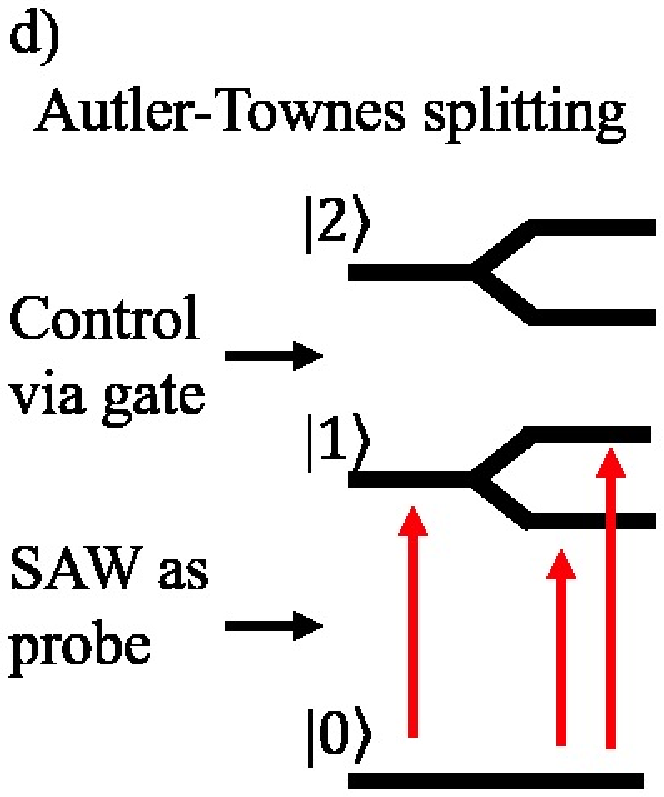} \label{fig:fig1d}
		\end{minipage}
	\end{minipage}
	\begin{minipage}[t]{0.49\textwidth}
		\vspace{0pt}
		\centering
		\includegraphics[width=0.9\textwidth]{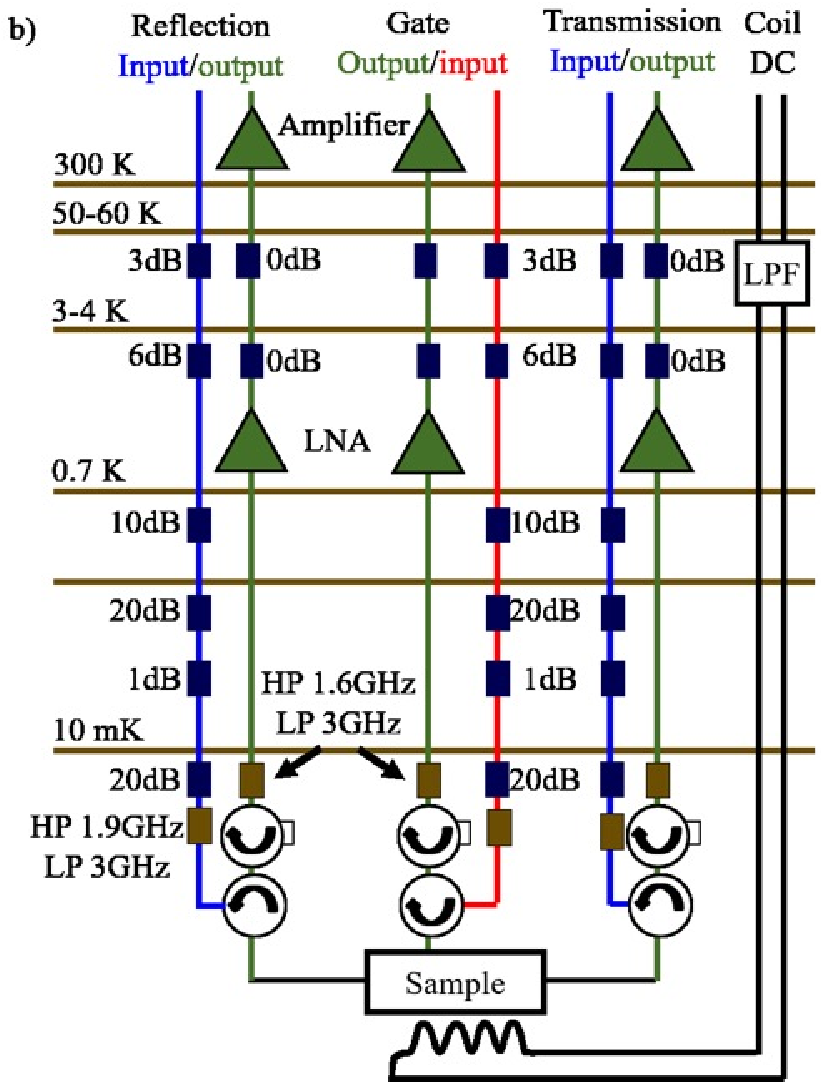}
		\label{fig:fig1b}
	\end{minipage}
\caption{(a) Layout of the measured sample. A transmon qubit is placed in between two IDTs. The left IDT (IDT A) is used for transduction to SAWs as well as detection of the reflected SAWs, and the right IDT (IDT B) is used for detection of the transmitted SAWs. The transmon qubit consists of a SQUID, a Qubit coupled IDT (QDT) and a capacitively coupled gate. (b) Schematic of the set-up for simultaneous transmission and reflection measurements. (c) Electron micrograph of a sample with the same design showing the the right IDT and the transmon qubit and the gate electrode. Both the QDT and the two IDTs are split finger transducers with a period of 1.28\,$\mu$m. The lower micrograph is a blow up of the SQUID and its connection to the QDT. (d) The transmittance of the transmon qubit is controlled by the Autler-Townes splitting. When a strong pulse is sent via the gate to the transmon at $f_{12}$, SAWs at $f_{01}$ can propagate between the IDTs without interacting with the qubit.}
\label{fig:Layout}
\end{figure*}

SAWs are mechanical waves that can propagate along the surface of any solid. They can be converted to and from electric microwave signals using interdigital transducers (IDTs) \cite{Smith1969} on piezoelectric substrates. In this work we have chosen gallium arsenide (GaAs) as the substrate. The IDT is an array of electrodes with alternating polarity, and when an ac-voltage is applied across the IDT, SAWs are generated traveling in both directions from the IDT. The spacing of the electrodes sets the centre frequency of the IDT ($f_{I}$) and the number of unit cells ($N_{p,I}$) determines the bandwidth by $BW_I=0.9f_{I}/N_{p,I}$.

An artificial atom interacting with SAWs can be designed using an IDT as a shunt capacitance for a superconducting quantum interference device (SQUID) to form a transmon qubit \cite{Koch2007}. This IDT will be referred to as the QDT (Qubit coupled IDT). Around the centre frequency of the QDT, $f_Q$, the transmon couples to SAWs with a coupling strength $\Gamma_{ac}=0.5f_Q K^2 N_{p,Q}$, where $K^2$ is the electromechanical coupling coefficient set by the substrate material and $N_{p,Q}$ is the number of unit cells in the QDT \cite{MVG,bookchapter}.

\section{Methods}
Our device consists of a transmon qubit, with a QDT as the shunting capacitance, positioned asymmetrically between two IDTs, see figure\,\ref{fig:Layout}a,c.  
The QDT as well as the IDTs are designed to couple to SAWs at 2.26\,GHz. The IDTs act as transducers converting microwave signals between the electromagnetic and acoustic domains. To reduce internal mechanical reflections both the IDTs and the QDT have split electrodes, where one period (unit cell) consists of two electrodes connected to the top bus bar and two to the bottom bus bar (figure\,\ref{fig:Layout}a,c)\cite{Bristol1972,Morgan,Datta}. Pertinent parameters are shown in table\,\ref{tab:sample2}. 

An optimized IDT should be impedance matched to 50\,$\Omega$, which can be achieved by varying the electrode overlap $W$ and $N_{p,I}$ \cite{Morgan,Datta}. However, impedance matching on GaAs requires a very large $N_{p,I}$ (approximately 450), and would result in a very small bandwidth of the IDTs. As a compromise, we have used $N_{p,I}=150$ to get a sufficient bandwidth of 14\,MHz, while achieving a reasonably good conversion efficiency. 
To obtain the IDT/QDT centre frequency, the electrodes have an equal spacing and width of $\lambda/8=158$\,nm. The electrode structure was made using electron beam lithography and a liftoff process with 27\,nm thick aluminium capped by 3\,nm of palladium. The transducers were connected to ground planes (5/85/10\,nm of Ti/Au/Pd) and the SQUID was deposited last using two-angle evaporation of 40/60\,nm aluminium at $\pm23^\circ$.

The acoustic transmission of the transmon qubit could be controlled in two different ways, which both operate by changing the energy level structure of the transmon such that the frequency of the 0-1 transition, $f_{01}$, is detuned with respect to the QDT frequency $f_Q$. The detuning is defined as $f_{01}-f_Q$ and the maximum transition frequency of the transmon is $f_{01,max}\approx (\sqrt{8E_JE_C}-E_C)/h\approx3.19$\,GHz \cite{Koch2007}.
i) Using the magnetic flux from an external magnetic coil, the energy levels of the transmon can be tuned over a wide range, this tuning is relatively slow. 
ii) The transmon can be driven by applying a microwave tone to the gate electrode and in particular the energy levels can be Rabi dressed creating an Autler-Towns splitting, see figure\,\ref{fig:Layout}d. The detuning achieved in this way is limited, but sufficiently large to detune the transmon from the SAW frequency. This type of control can also be very fast. 



The device was measured using the set-up shown in figure\,\ref{fig:Layout}b. When the SAWs are generated by IDT A, the SAWs propagate with the amplitude $\phi_R^+$ towards the transmon. At the transmon the SAWs are reflected ($\phi_L^-$) back towards IDT A or transmitted ($\phi_R^-$) towards IDT B, where the reflected and transmitted SAWs are measured simultaneously. Here, the indices L/R refer to left and right respectively.  The reflection coefficient of the transmon is then given by $r=\phi_L^-/\phi_L^+$ and the transmission coefficient by $t=\phi_R^-/\phi_L^+$. The coefficients $t$ and $r$ are related by the definition $t=r+1$. 

The experiment is operated at a temperature of 12\,mK in order to be in the quantum limit, $k_BT\ll \hbar \omega$. 
In transmission, a continuous signal was sent to IDT A where it was partly electrically reflected due to impedance mismatch and partly transduced into SAWs in both directions and the right propagating waves were measured at IDT B. The measured signal in the frequency domain has been Fourier-filtered
to remove multiple acoustic reflections and electric crosstalk \cite{Campbell1989,Ekstrom2017}. 

\section{Results}
\begin{figure*}
\centering
\includegraphics[width=0.49\linewidth]{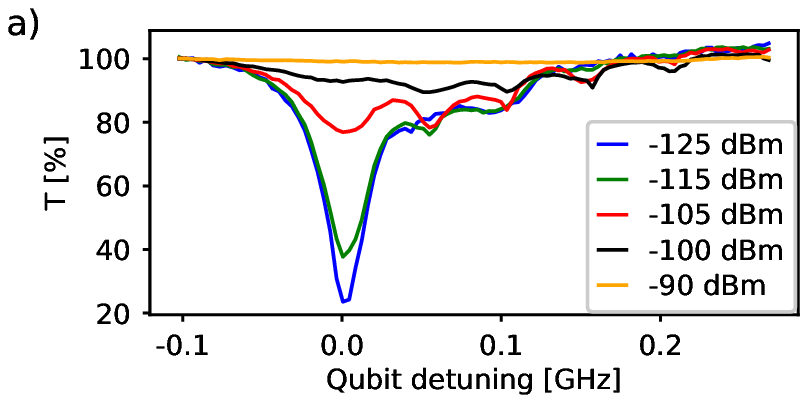}
\includegraphics[width=0.49\linewidth]{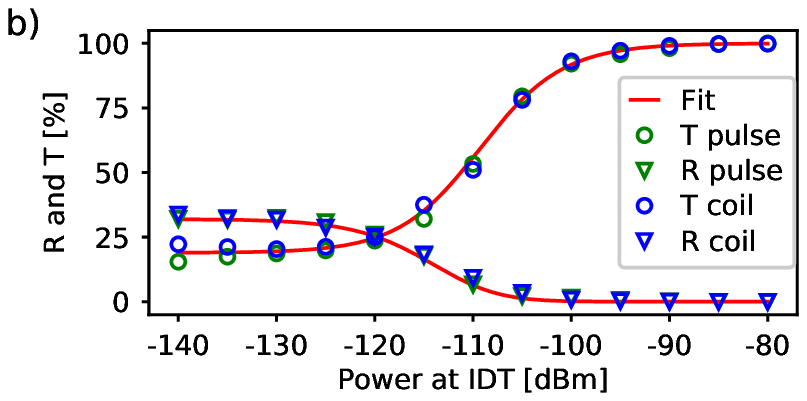}
\caption{\label{fig:Characterization} (a) Acoustic transmission between IDT A and B while changing the flux through the SQUID loop and changing the power of the signal sent to IDT A. On resonance with the SAWs, the qubit blocks the transmission. At high powers several  transitions of the transmon are visible. (b) Difference between the on and off resonant reflection (R, triangles) and  transmission (T, circles) measured simultaneously at the center frequency of the IDT versus input power to IDT A. The transmon was either tuned on or off resonance with the SAWs using the external coil (blue) or by pulsing the gate signal that controlled the Autler-Townes splitting (green). At low powers 80\,\% extinction is observed in the transmitted field. The fit to both measurements (red lines) gives a pure dephasing of $\Gamma_{\phi}$=8\,MHz and a decay rate which is determined by the acoustic coupling $\Gamma_{ac}=$21\,MHz, in good agreement with the designed value.}
\end{figure*}

We measure the acoustic scattering of the transmon for different SAW power while tuning the transmon frequency (see figure\,\ref{fig:Characterization}). The transmittance is given by $T=|t|^2$, similarly the reflectance is given by $R=|r|^2$. 

\subsection{Controling SAW propagation with magnetic flux}
We start by controling the transmon with an external magnetic flux from a coil (figure\,\ref{fig:Characterization}a).
SAWs are launched from IDT A and the transmitted SAWs are measured using IDT B. When the transmon is tuned out of resonance with the frequency of the SAWs, $f_I$, the SAWs are transmitted without interacting with the transmon. On resonance, the SAWs are partly reflected resulting in a reduction of the transmission at zero detuning. As the SAW power is increased the qubit is gradually saturated and the transmittance approaches unity.
At small positive detuning and increasing power to IDT A, higher energy transitions corresponding to $f_{02}/2$ and $f_{03}/3$ can be observed. These features are in good agreement with the estimated anharmonicity of $f_{12}-f_{01}=-129$\,MHz. 

How the transmittance depends on SAW power is shown in figure\,\ref{fig:Characterization}b, the transmittance (blue triangles) and reflectance (blue dots) at zero detuning were measured simultaneously. We see that the transmittance increases with increasing SAW power whereas the reflectance decreases.

When the 0-1 transition frequency of the transmon is on resonance with the frequency of the QDT, $\it{i.e.}$
at zero detuning, the reflection coefficient is given by \cite{Astafiev2010}

\begin{equation}
r=-\frac{r_0}{1+\Omega_p^2/\Gamma_{01}\gamma_{01}}.
\label{eq:reflection}
\end{equation}

Here $r_0=1/(1+2\Gamma_\phi/\Gamma_{01})$ is the maximum reflection amplitude, $\Gamma_{01}\approx\Gamma_{ac}$ is the relaxation rate from $|1\rangle$ to $|0\rangle$ of the transmon, $\gamma_{01}=\Gamma_{01}/2+\Gamma_\phi$ is the 0-1 decoherence rate, $\Gamma_\phi$ is the 0-1 pure dephasing rate and $\Omega_p$ is the Rabi frequency at which the coherent incoming SAWs drive coherent oscillations of the transmon. The Rabi frequency is proportional to the amplitude of the SAW field, $\Omega_p=k \sqrt{P}$, where  $P$ is the incoming SAW power. 

On resonance, the incoming SAWs interfere constructively with the reflected field from the transmon and destructively with the transmitted field. This should result in perfect reflection and zero transmission on resonance, in the absence of pure dephasing and if the power of the incoming SAWs is low ($\Omega_p\ll\Gamma_{10},\gamma_{10}$) \cite{Shen2005,Shen2005PRL}. From figure\,\ref{fig:Characterization}b it is clear that the pure dephasing is not negligible in this device.

In figure\,\ref{fig:Characterization}b the measured reflection and transmission is fitted to the reflectance and the transmittance as a function of $P$ using $k$ and $\Gamma_\phi$ as the only fit parameters. Both fits are in good agreement with an acoustic coupling of $21$\,MHz, where $\Gamma_{01}=\Gamma_{ac}$. At low powers the deviation from zero in transmission and from $100\%$ in reflection is due to a pure dephasing $\Gamma_\phi \approx 8$\,MHz. The pure dephasing limits the extinction in transmission at low SAW powers to $80\%$. 

\subsection{Controlling SAW propagation with the Autler-Townes splitting}
\begin{figure*}[t]
\includegraphics[width=0.32\linewidth]{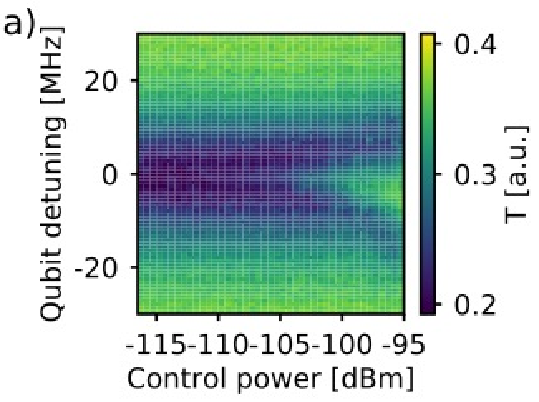}
\includegraphics[width=0.32\linewidth]{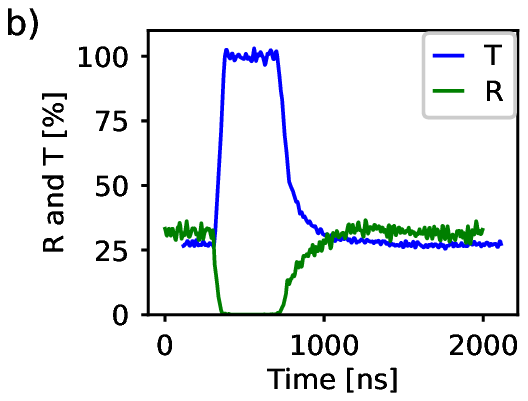}
\includegraphics[width=0.32\linewidth]{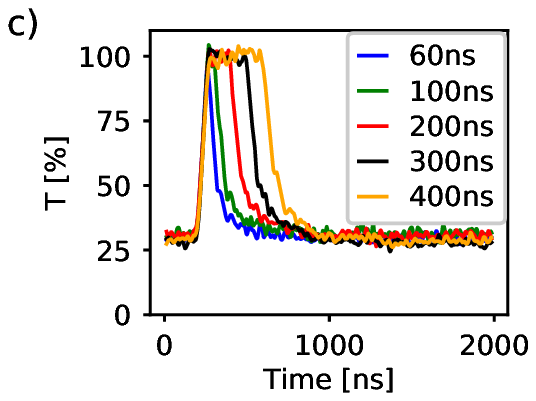}
\caption{\label{fig:Router} (a) Acoustic transmission of weak SAWs at $f_{01}$ while increasing the power of the control signal at $f_{12}$ (sent via the gate) shows an increasing Autler-Townes splitting. (b) The Autler-Townes splitting induced by a control pulse at $f_{12}$ is used to change the frequency of the transmon such that it is out of resonance with the SAWs while the pulse is on. When the pulse is on the SAWs are transmitted (blue) instead of reflected (green). (c) The transmission amplitude does not change when the pulse length of the control is decreased down to 60\,ns.} 
\end{figure*}

Hitherto the 0-1 transition frequency of the transmon has been tuned with an external coil, but it can also be changed based on Autler-Townes splitting as shown in figure\,\ref{fig:Layout}d \cite{AutlerTownes1955,Hoi2011}. The levels $\left|1\right>$ and $\left|2\right>$ are dressed by the control field, causing an Autler-Towns splitting of the 0-1 transition. Using two-tone spectroscopy, we apply a weak continuous SAW signal via IDT A at $f_{01}$ as a probe. At the same time, we apply a control signal via the capacitively coupled gate (see figure\,\ref{fig:Layout}) at the 1-2 transition frequency, $f_{12}$,
and measure both reflection (IDT A) and transmission (IDT B) at $f_{01}$. For no or weak control powers, the SAWs are partly reflected, and we observe 80\% extinction as described above and shown in figure\,\ref{fig:Characterization}b. However, for substantial control powers the energy levels $\left|1\right>$ and $\left|2\right>$ Autler-Townes split \cite{Abdumalikov2010}. This results in two transition frequencies different from $f_{01}$, \textit{i.e.} the transmon is no longer on resonance with the frequency of the SAWs and the SAWs are transmitted instead of being reflected. The Autler-Townes splitting can be seen in figure\,\ref{fig:Router}a, where the transmission between IDT A and IDT B is plotted versus qubit detuning.

As a control experiment we again probe the reflected and transmitted fields versus power to IDT A, but now use the Autler-Townes splitting to change the transmon in or out of resonance with the SAWs. The measured reflectance (green triangles in figure\,\ref{fig:Characterization}b) and transmittance (green dots) was fitted to \,(\ref{eq:reflection}). This gave very similar results as the measurements performed using the external coil. In figure\,\ref{fig:Characterization}b the fit shown (red lines) is the combined fit to both measurements.

\subsection{Fast control of the SAW propagation}
The control of SAW propagation be exploited to create a phonon-router in the quantum regime, similar to the photon router that was demonstrated in \,\cite{Hoi2011}. In principle either flux control or Autler-Towns control could be used. Here we show a proof of principle of this concept using the Autler-Towns control since it allows fast changes of the transmittance.

Initially the frequency of the qubit, $f_{01}$, is is tuned to be on resonance with the SAWs. A strong control signal is then applied as a pulse to the qubit gate (see figure\,\ref{fig:Layout}d). When the pulse is OFF, the SAWs are partly reflected by the transmon and when it is ON, the SAWs are transmitted. This is shown in figure\,\ref{fig:Router}b for reflection (green) and transmission (blue) versus time when a $400$\,ns pulse is applied as the control. The rise time of the router is about 40 ns, limited by the bandwidth of the IDT, while the fall time of the router is $160$\,ns. The longer fall time is due to multiple transits between IDT B and the transmon. Routing using shorter control pulses is shown in figure\,\ref{fig:Router}c, where the amplitude of the transmitted SAW field is the same down to $60$\,ns control pulses. Detection of shorter control pulses is limited by the bandwidth of the IDTs. 

\subsection{Manipulating SAW pulses}
\begin{figure*}
\includegraphics[width=0.49\linewidth]{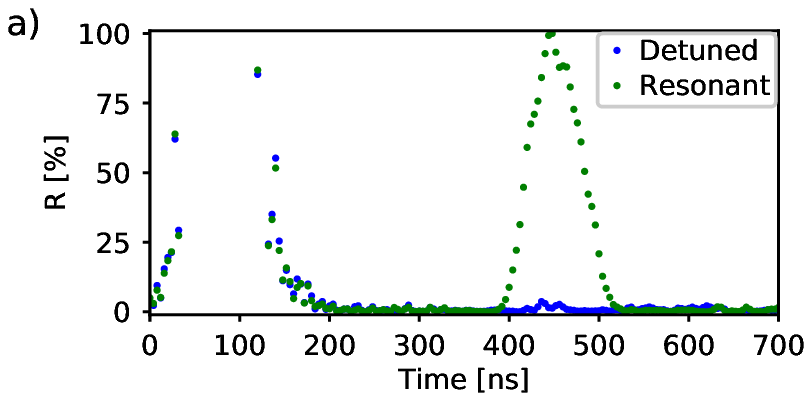}
\includegraphics[width=0.49\linewidth]{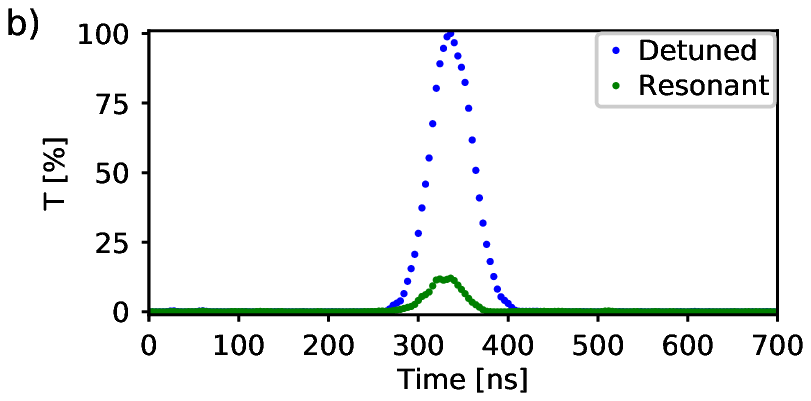}
\caption{\label{fig:Pulsing} (a) Reflectance and (b) transmittance of 100\,ns SAW pulses when the transmon is on resonance with the SAWs (green) and detuned from the frequency of the SAWs (blue) using the external magnetic coil. In the reflection there is a large electric reflection due to the impedance mismatch between the IDT and 50\,$\Omega$. The reflection is normalized to the maximum of the on resonance acoustic reflection observed at a time around 450\,ns. The transmission is normalized to the maximum transmission, which occurs when the transmon is detuned. }
\end{figure*}

For more advanced in-flight manipulation of SAWs, it will be important to not only route continuous SAWs but also route SAW pulses. The possibility to do this with our device is shown in figure\,\ref{fig:Pulsing}, where a $100$\,ns SAW pulse generated at IDT A is routed using the external coil to flux tune the transmon in and out of resonance with the SAWs. When the transmon is out of resonance with the SAWs (blue), the SAW pulse is transmitted without interacting with the transmon. On resonance (green), the SAW pulse is mostly reflected but part of it is also transmitted even at low powers of the SAW pulse. This is due to incoherent scattering, which in turn is due to the pure dephasing as mentioned previously. In this measurement, imperfect reflection is due to pure dephasing, but even without pure dephasing the reflection would be limited by the relaxation rate of the transmon in comparison to the length of the SAW pulse. The pulse length has to be longer than the relaxation time in order to reach full reflection, which will be important to consider for future experiments. 

\section{Discussion and conclusion}
Here we have shown that routing SAWs at the phonon level is feasible. The performance of the phonon router presented here is limited by the pure dephasing which in turn limits the maximum reflection from the qubit. We have also shown that the routing can be done very fast, on the 100\,ns timescale. In comparison to the photon router in\,\cite{Hoi2011}, the phonon router presented here has a four times longer rise time and the pulse length is limited to six times longer due to the narrow bandwidth of the IDTs. This proof of principle demonstration is promising for further in-flight manipulation of propagating single phonons.

\begin{table*}
\centering
\caption{\label{tab:sample2} Table of material parameters, design and measured/extracted values for the measured sample. The dielectric constant, $\epsilon_\infty$, the sound velocity, $v_0$, and electromechanical coupling constant, $K^2$ are given by the material which is GaAs. The parameters that are set by design are: IDT A to qubit distance, $d_{Q,A}$, IDT B to qubit distance, $d_{Q,A}$, number of finger pairs of the IDT, $N_{p,I}$, and the QDT, $N_{p,Q}$, bandwidth of the IDT, $BW_I$, and the QDT, $BW_Q$, finger overlap (IDT/QDT width), $W$, and SQUID area $A_{SQ}$. The measured or extracted parameters are, maximum qubit frequency, $f_{01,max}$, IDT/QDT frequency, $f_{I/Q}$, Josephson coupling energy at zero flux $E_{J,0}$, Charging energy, $E_C$, $E_J/E_C$ ratio, acoustic coupling, $\Gamma_{ac}$, and SAW wavelength, $\lambda_0$}
\resizebox{\textwidth}{!}{
\begin{tabular}{ c | c  c  c | c  c  c  c  c  c  c  c | c  c  c  c  c  c  c }
\br
& \multicolumn{3}{ c |}{Material} & \multicolumn{8}{ c |}{Design} & \multicolumn{7}{ c }{Measured}\\
\mr
Parameter & $\epsilon_\infty$ & $v_0$ & $K^2$ & $d_{Q,A}$ & $d_{Q,B}$ & $N_{p,I}$ & $N_{p,Q}$ & $BW_I$ & $BW_Q$ & $W$ & $A_{SQ}$ & $f_{01,max}$ & $f_I$=$f_Q$ & $E_{J,0}$ & $E_C$ & $E_J/E_C$ & $\Gamma_{ac}$ & $\lambda_0$\\
Value & 12.0			 &  2864 &  0.07    &  300	          &  100	     &  150	         &  25	     &  14	     &  81	       &  40	  &  16.5             &  3.19	          &  2.2641     &  10.7                  &  129     &  83         &  21	              &  1265\\
Unit 	&				& m/s     & $\%$    & $\mu$m	  & $\mu$m   &			&		     & MHz	     & MHz	       & $\mu$m & $\mu$m$^2$ & GHz	 & GHz	     & GHz	                  & MHz	&	         & MHz                   & nm\\	
\br
\end{tabular}
}
\end{table*}

An interesting development of the phonon router presented here would be to modify the device to include two transmons between the two IDTs, where the transition frequencies of the two transmons can be tuned individually either with separate capacitively coupled gates or flux lines. If the first transmon is tuned out of resonance and the second transmon on resonance with the frequency of the SAWs, a SAW pulse generated at IDT~A can propagate through the first transmon and be reflected at the second transmon. Since it takes a substantial time for the SAWs to propagate, there is ample time to tune the first transmon on resonance with the SAWs and in this way trap the SAW pulse between the two transmons. If the transmons have low pure dephasing, the SAW pulse should be able to travel back and fourth without too much dissipation since SAWs propagate the substrate with low propagation losses \cite{bookchapter,Ekstrom2017}. The trapped SAW pulse can then be released controllably to the left or to the right by either tuning the first transmon or the second transmon out of resonance with the SAWs.

In conclusion, we have demonstrated that a superconducting transmon qubit, designed to interact with surface acoustic waves, can be used to route propagating phonons. On resonance with the incident SAWs and at low powers, the transmon reflects the SAW field and an extinction of 80\,\% was observed in transmission. Using Autler-Townes splitting of the transmon's transition frequency the reflection and transmission of the SAW field could also be controlled in time and the propagating phonons could be routed with a rise time of $40$\,ns. Moreover, we showed that it is possible to route short (100 ns) pulses, which further enables experiments where an acoustic pulse is captured between two transmons and released in a controlled way. In such experiments it is beneficial to use SAW phonons since they have a five orders of magnitude slower speed than photons in vacuum allowing for more time to perform in-flight manipulations of the propagating phonons.

\ack
We wish to acknowledge financial support from The Knut and Alice Wallenberg Foundation and The Swedish Research Council. The samples were made at the Nanofabrication Laboratory at Chalmers University of Technology. We thank Marcus Rommel for assistance imaging the sample and we acknowledge fruitful discussions with Sankar Sathyamoorthy, Martin Gustafsson and Anton Frisk-Kockum. 


\section*{References}
\bibliography{bib_phononRouter}

\end{document}